\documentclass{jfm}

\usepackage{silence} 
\usepackage{preamble}
\graphicspath{{./Figures/}}

\DeclareMathOperator{\Sk}{Sk}
\DeclareMathOperator{\As}{As}

\newcommand{\fourier}{\mathcal{F}}
\newcommand{\hilbert}{\mathcal{H}}

\renewcommand*{\epsilon}{\varepsilon}

\shorttitle{Wind-induced wave shape changes in shallow water}
\shortauthor{T.\ Zdyrski  and F.\ Feddersen}

\title{Wind-induced changes to surface gravity wave shape in shallow water}

\author{Thomas \href{https://orcid.org/0000-0003-3039-172X}{Zdyrski}\aff{1}
  \corresp{\email{tzdyrski@uscd.edu}},
  \and Falk \href{https://orcid.org/0000-0002-5488-9074}{Feddersen}\aff{1}}

\affiliation{\aff{1}Scripps Institution of Oceanography, UCSD, La Jolla, CA 92092-0209, USA}

\begin{document}

\maketitle

\begin{abstract}
Wave shape (\eg{} wave skewness and asymmetry) impacts sediment
transport, remote sensing and ship safety.
Previous work showed that wind affects wave shape in intermediate and
deep water.
Here, we investigate the effect of wind on wave shape in shallow water
through a wind-induced surface pressure for different wind speeds and
directions to provide the first theoretical description of wind-induced
shape changes.
A multiple-scale analysis of long waves propagating over a shallow,
flat bottom and forced by a Jeffreys-type surface pressure yields a
forward or backward Korteweg--de Vries (KdV)--Burgers equation for the
wave profile, depending on the wind direction.
The evolution of a symmetric, solitary-wave initial condition is
calculated numerically.
The resulting wave grows (decays) for onshore (offshore) wind and
becomes asymmetric, with the rear face showing the largest shape
changes.
The wave profile's deviation from a reference solitary wave is primarily
a bound wave and trailing, dispersive, decaying tail.
The onshore wind increases the wave's energy and skewness with time
while decreasing the wave's asymmetry, with the opposite holding for
offshore wind.
The corresponding wind speeds are shown to be physically realistic, and
the shape changes are explained as slow growth followed by rapid
evolution according to the unforced KdV equation.
\end{abstract}


\section{\label{sec:introduction} Introduction}
\noindent
The study of wind and ocean wave interactions began with
\citet{jeffreys1925formation} and continues to be an active field of
research~\citep[\eg][]{janssen1991quasi,donelan2006wave,sulivan2010dynamics}.
Many theoretical
studies~\citep[\eg][]{jeffreys1925formation,miles1957generation,phillips1957generation}
focus on calculating wind-induced growth rates and often employ
phase-averaging techniques.
However,
experimental~\citep[\eg][]{leykin1995asymmetry,feddersen2005wind} and
theoretical~\citep[\eg][]{zdyrski2020wind} studies have shown wind can
also influence wave shape, quantified by third-order shape statistics
such as skewness and asymmetry, corresponding to vertical and horizontal
asymmetry, respectively.
Furthermore, while many numerical studies on coupled wind and waves
employ sinusoidal water waves and therefore neglect wind-induced shape
changes~\citep[\eg][]{hara2015wave,husain2019boundary}, some recent
numerical studies have incorporated wind-induced changes to the
wave field using coupled air--water
simulations~\citep[\eg][]{liu2010numerical,hao2019wind} or direct
numerical simulations of two-fluid
flows~\citep[\eg][]{zonta2015growth,yang2018direct}.
Wave shape influences sediment transport, affecting beach
morphodynamics~\citep[\eg][]{drake2001discrete,hoefel2003wave}, while
wave skewness affects radar altimetry
signals~\citep[\eg][]{hayne1980radar} and asymmetry influences ship
responses to wave impacts~\citep[\eg][]{soares2008abnormal}.

Waves in shallow water, where $kh \ll 1$ (with $h$ the water depth,
$k=2\pi/\lambda$ the wavenumber and $\lambda$ the wavelength), differ
qualitatively from those in intermediate ($kh \sim 1$) to deep ($kh \gg
1$) water.
For waves with small amplitudes $a_0 \ll h$,
leveraging the small parameters $a_0/h \sim (kh)^2 \ll 1$
yields the Boussinesq equations with weak dispersion and nonlinearity.
When dispersion balances nonlinear
focusing, a special class of waves, known as solitary waves, are formed
and appear in environments ranging from nonlinear optical pulses
\citep[\eg][]{kivshar1993dark} to astrophysical dusty plasmas
\citep[\eg][]{sahu2012nonextensive}.
These well-understood waves are often used to study fluid
dynamical~\citep[\eg][]{munk1949solitary,hammack1974korteweg,miles1979korteweg,lin1998numerical}
and
engineering~\citep[\eg][]{monaghan1999solitary,lin2004numerical,xu2018prediction}
contexts owing to their simplicity.
One of the simplest equations displaying solitary waves is the
Korteweg--de Vries (KdV) equation, which incorporates dispersion and
nonlinearity.
When augmented with a dissipative term, this becomes the KdV--Burgers
equation, with applications to damped internal tides
\citep[\eg][]{sandstrom1995dissipation}, electron waves in graphene
\citep[\eg][]{zdyrski2019effects} and viscous flow in blood vessels
\citep[\eg][]{antar1999weakly}.
While field observations~\citep[\eg][]{cavaleri1981wind} have
investigated the wind-induced growth of shallow-water waves, the
interaction of wind and shallow-water waves has not yet been formulated
into a simple equation such as the KdV--Burgers equation.

The influence of wind on wave shape has been previously investigated in
intermediate and deep water~\citep{zdyrski2020wind}.
However, the coupling between wind and wave shape has not yet been
investigated in shallow water.
To investigate wind and surface wave interactions in shallow water over
a flat bottom, we introduce a wind-induced pressure term to the
Boussinesq equations in \cref{sec:derivation}.
The resulting KdV--Burgers equation governs a solitary wave's evolution,
which we solve numerically
to yield the wave's energy, skewness and asymmetry
in \cref{sec:results}.
We calculate the wind speed, discuss the asymmetry and
compare our results to intermediate- and deep-water waves in
\cref{sec:discussion}.

\section{\label{sec:derivation} Derivation of the KdV--Burgers equation}

\subsection{\label{sec:governing} Governing equations}
\noindent
We treat the flow as irrotational and inviscid and neglect surface
tension.
Furthermore, we restrict ourselves to planar wave propagation in the
$+x$ direction.
Finally, we choose a coordinate system with $z=0$ at the mean water
level and a horizontal, flat bottom located at $z=-h$.
Then, the incompressibility condition and standard boundary conditions
are
\begin{alignat}{2}
  0 &= \phi_{xx} + \phi_{zz} &&\qq{on}
  -h < z < \eta , \label{eq:laplace}\\
  0 &= \phi_{z} &&\qq{on} z=-h , \label{eq:bottom_bc}\\
  \phi_{z} &= \eta_{t} + \phi_{x} \eta_{x} &&\qq{on} z = \eta ,
  \label{eq:kinematic_bc}\\
  0 &= \frac{p}{\rho_w} + g\eta + \phi_{t} +
  \frac{1}{2} \bqty{\phi_{x}^2 + \phi_{z}^2} &&\qq{on} z=
  \eta . \label{eq:dynamic_bc}
\end{alignat}
Here, $\eta(x,t)$ is the wave profile, $\phi(x,z,t)$ is the flow's
velocity potential related to the velocity $\vec{u} = \grad{\phi}$,
$p(x,t)$ is the surface pressure, $g$ is the gravitational acceleration
and $\rho_w$ is the water density.
We used the $\phi$ gauge freedom to absorb the Bernoulli `constant'
$C(t)$ in the dynamic boundary condition.
We seek a solitary, progressive wave which decays at infinity,
$\eta(\vec{x},t) \to 0$ as $\abs{\vec{x}} \to \infty$, with similar
conditions on $\vec{u}$.
We choose a coordinate system where the average bottom horizontal
velocity vanishes,
\begin{equation}
  \overline{ \pdv{\phi}{x} } = 0 \qq{on} z=-h ,
  \label{eq:bot_bc_horz}
\end{equation}
with the overline a spatial average $\overline{f} \coloneqq
\lim_{L\to\infty} \int_{-L}^{L} f \dd{x} / (2L)$.
Additionally, we assume the surface pressure $p(x,t)$ is a Jeffreys-type
forcing \citep{jeffreys1925formation},
\begin{equation}
  p(x,t) = P \pdv{\eta(x,t)}{x} .
  \label{eq:press_def}
\end{equation}
Here, $P$ is proportional to $(U-c)^2$, with $c$ the wave's nonlinear
phase speed and $U$ the wind speed (\cf{} \cref{sec:press_mag}).
Note that $P>0$ corresponds to (`onshore') wind in the same direction as
the wave while $P<0$ denotes (`offshore') wind opposite the wave. We use
a Jeffreys forcing for its analytic simplicity and clear demonstration
of wind--wave coupling.
Jeffrey's separated sheltering mechanism is likely only relevant
in special situations (\eg{} near breaking,
\citealp{banner1976separation}, or for steep waves under strong winds,
\citealp{touboul2006interaction,tian2013evolution}).
Additionally, numerical simulations of sinusoidal waves suggest the peak
surface pressure is shifted approximately \SI{135}{\degree} from the
wave peak, while Jeffreys would give a \SI{90}{\degree}
shift~\citep{husain2019boundary}.
However, a fully dynamic coupling between wind and waves --- necessary
for an accurate surface pressure over a non-sinusoidal, dynamic water
surface --- is outside the scope of this paper.
Furthermore, the applicability of Jeffreys forcing to extreme waves
means our theory could apply to the wind forcing of rogue waves in
shallow water~\citep{kharif2008influence}.

\subsection{\label{sec:nondim} Non-dimensionalization}
\noindent
We non-dimensionalize our system with the known characteristic
scales: the horizontal length scale $L$ over which $\eta$ changes
rapidly, expressed as an effective wavenumber $k_E \coloneqq 2 \pi/L$;
the (initial) wave amplitude $a_0 = H_0/2$ (\ie{} half the wave height
$H_0$); the depth $h$; the gravitational acceleration $g$; and the wind
speed $U$,
expressed as a pressure magnitude $P \propto \rho_a (U-c)^2$,
with $\rho_a \approx \num{1.225e-3} \rho_w$ the density of air.
Denoting non-dimensional variables with primes, we have
\begin{equation}
  \begin{aligned}
  x &= \frac{x'}{k_E} = h \frac{x'}{\sqrt{\mu_E}}, \\
  z &= h z' ,
  \end{aligned}
  \qquad
  \begin{aligned}
  t &= \frac{t'}{k_E c_0}
    = \frac{t'}{\sqrt{\mu_E}} \sqrt{\frac{h}{g}} , \\
  P &= \epsilon P' \frac{\rho_w g}{k_E}
    = \frac{\epsilon}{\sqrt{\mu_E}} P' \rho_w c_0^2 ,
  \end{aligned}
  \qquad
  \begin{aligned}
  \eta &= a_0 \eta' = h \epsilon \eta' , \\
  \phi &= \phi'\frac{a_0}{k_E}\sqrt{\frac{g}{h}}
    = \frac{\phi'\epsilon}{\sqrt{\mu_E}} c_0 h ,
  \end{aligned}
  \label{eq:nondim_expressions}
\end{equation}
with linear, shallow-water phase speed $c_0 = \sqrt{g h}$.
Our system's dynamics is controlled by three small, non-dimensional
parameters: $\epsilon \coloneqq a_0/h$, $\mu_E \coloneqq (k_E h)^2$ and
$P k_E/(\rho_w g)$.
We will later require $\order{\epsilon} = \order{\mu_E} =
\order{Pk_E/(\rho_w g)}$.
Now, our non-dimensional equations take the form
\begin{alignat}{2}
  0 &= \mu_E \phi'_{x'x'} + \phi'_{z'z'} &&\qq{on}
    -1 < z' < \epsilon \eta' , \label{eq:laplace_nondim} \\
  0 &= \phi'_{z'} &&\qq{on} z'=-1 , \label{eq:bottom_bc_nondim} \\
  \phi'_{z'} &= \mu_E \eta'_{t'} +
    \epsilon \mu_E \phi'_{x'} \eta'_{x'} &&\qq{on} z' = \epsilon \eta' ,
    \label{eq:kinematic_bc_nondim} \\
  0 &= \epsilon P' \eta'_{x'} +  \eta' + \phi'_{t'} + \frac{1}{2}
    \pqty{\epsilon \phi_{x'}^{\prime .2} + \frac{\epsilon}{\mu_E}
    \phi_{z'}^{\prime .2}} &&\qq{on} z'= \epsilon \eta' .
    \label{eq:dynamic_bc_nondim}
\end{alignat}
We will drop the primes throughout the remainder of this section for readability.

\subsection{\label{sec:boussinesq} Boussinesq equations, multiple-scale
expansion, KdV equation and initial condition}
\noindent
Here, we modify the Boussinesq equation's derivation provided by
\citet{mei2005nonlinear} or \citet{ablowitz2011nonlinear} by including
the surface pressure forcing in \cref{eq:dynamic_bc}.
Taylor expanding the velocity potential $\phi$ about the bottom,
$z=-1$, and applying Laplace's
equation \cref{eq:laplace_nondim} and the bottom boundary
condition \cref{eq:bottom_bc_nondim} yields an expansion of $\phi$ in
terms of $\mu_E \ll 1$ and the velocity potential at the bottom,
$\varphi \coloneqq \eval{\phi}_{z=-1}$.
This $\phi$ expansion can be substituted into the two remaining boundary
equations, \cref{eq:kinematic_bc_nondim,eq:dynamic_bc_nondim}, to give
the Boussinesq equations with a pressure forcing term,
\begin{gather}
  \partial_t \eta + \partial_x^2 \varphi + \epsilon \partial_x
    \pqty{\eta \partial_x \varphi} -\frac{1}{6}\mu_E \partial^4_x
    \varphi = \order{\mu_E^2} \label{eq:kinematic_bc_varphi} , \\
  \partial_t \varphi + \epsilon P \partial_x \eta + \eta -
    \frac{1}{2}\mu_E \partial_t \partial_x^2 \varphi +
    \frac{1}{2}\epsilon\pqty{\partial_x \varphi}^2 = \order{\mu_E^2} .
    \label{eq:dynamic_bc_varphi}
\end{gather}
Further, we will now assume $\order{\epsilon} = \order{\mu_E} \ll 1$.

We now expand $t$ using multiple time scales $t_n =
\epsilon^n t$ for $n= 0,1$, so all time derivatives become $\partial_t \to
\partial_{t_0} + \epsilon \partial_{t_1}$.
Then, we write $\eta$ and $\varphi$ as asymptotic series of $\epsilon$,
\begin{equation}
  \eta(x,t) = \sum_{k=0}^{\infty} \epsilon^k
    \eta_{k}(x,t_0,t_1) \qq{and}
  \varphi(x,t) = \sum_{k=0}^{\infty} \epsilon^k
    \varphi_{k}(x,t_0,t_1,) .
\end{equation}
Now, we will reduce the Boussinesq equations,
\cref{eq:kinematic_bc_varphi,eq:dynamic_bc_varphi}, to the KdV equation
following a similar method to \citet{mei2005nonlinear} and
\citet{ablowitz2011nonlinear}.
Collecting order-one terms $\order{\epsilon^0}$ from
\cref{eq:kinematic_bc_varphi,eq:dynamic_bc_varphi} gives
a wave equation for $\eta_0$ and $\phi_0$.
The right-moving solutions are
\begin{equation}
  \varphi_0 = f_0(x-t_0,t_1) \qq{and}
  \eta_0 = f'_0(x-t_0,1) \qq{with}
  f_0' \coloneqq \eval{\pdv{f_0(\theta,t_1)}{\theta}}_{\theta = x-t_0} .
\end{equation}
Continuing to the next order of perturbation theory, we retain terms of
$\order{\epsilon}$,
\begin{gather}
    \pdv{\eta_1}{t_0} + \pdv[2]{\varphi_{1}}{x} =
      -\pdv{\eta_0}{t_1} - \pdv{x} \pqty{\eta_0 \pdv{\varphi_0}{x}} +
      \frac{1}{6} \frac{\mu_E}{\epsilon} \pdv[4]{\varphi_0}{x} ,
  \\
    \eta_1 + \pdv{\varphi_1}{t_0} = -P \pdv{\eta_0}{x} -\pdv{\varphi_0}{t_1}
      + \frac{1}{2} \frac{\mu_E}{\epsilon} \frac{\partial^3 \varphi_0}
        {\partial t_0 \partial^2 x}
      - \frac{1}{2} \pqty{ \pdv{\varphi_0}{x} }^2
  .
\end{gather}
Inserting our leading-order solutions for $\eta_0$ and $\varphi_0$,
eliminating $\eta_1$ and preventing resonant forcing of $\varphi_1$
gives the KdV--Burgers equation,
\begin{equation}
  \pdv{\eta_0}{t_1} + \frac{3}{2}
    \eta_0 \pdv{\eta_0}{x} + \frac{1}{6} \frac{\mu_E}{\epsilon}
    \pdv[3]{\eta_0}{x} = -P \frac{1}{2} \pdv[2]{\eta_0}{x} .
  \label{eq:kdv_burgers}
\end{equation}
Note that \cref{eq:kdv_burgers} has a rescaling symmetry, with $\mu_E
\to \lambda^2 \mu_E$ equivalent to taking $(x,t_0,t_1,P) \to
(x,t_0,t_1,P)/\lambda$.
Therefore, we fix the length scale (equivalently, $k_E$) by choosing
$\mu_E = 6 \epsilon$.
Note that incorporating slowly varying bottom bathymetry $\partial_x h =
\order{\epsilon}$ can yield an equation of the form
\cref{eq:kdv_burgers} with spatially varying
coefficients~\citep[\eg][]{johnson1972some,ono1972wave}, although such
an analysis is outside the scope of this study.

For offshore wind, the pressure term $P \partial^2_x \eta_0$ acts as a
positive viscosity causing damping, and \cref{eq:kdv_burgers} is the
(forward) KdV--Burgers equation with $P<0$.
However, for onshore wind, the viscosity is negative and causes wave
growth, yielding the backward KdV--Burgers equation with $P>0$.
The backward KdV--Burgers equation is ill posed in the sense of Hadamard
because the solution is highly sensitive to changes in the initial
condition~\citep{hadamard1902problemes}.
While a finite-time singularity (\ie{} wave breaking) is likely, the
multiple-scale expansion used to derive \cref{eq:kdv_burgers} is only
valid for time intervals of $\order{1/\epsilon}$, and we limit our
analysis to short times removing the need to regularize the solution.

The solitary-wave solutions of the unforced ($P=0$) KdV equation exist
due to a balance of dispersion $\partial_x^3 \eta_0$ with focusing
nonlinearity $\eta_0 \partial_x \eta_0$ and have the
form~\citep[\eg][]{mei2005nonlinear}
\begin{equation}
  \eta_0 = H_0 \sech^2\pqty{\frac{x}{\Delta}}
  \qq{with}
  \Delta = \sqrt{\frac{8}{H_0}} ,
  \label{eq:initial_condition}
\end{equation}
in a co-moving frame with $H_0>0$ an order-one parameter.
For reference, unforced solitary waves travel relative to the laboratory
frame with non-dimensional, nonlinear phase
speed~\citep[\eg][]{mei2005nonlinear}
\begin{equation}
  c = 1 + \epsilon \frac{H_0}{2}
  \label{eq:nonlin_phase_speed}
\end{equation}
We use \cref{eq:initial_condition} for our initial condition and choose
$H_0 = 2$ so the initial, dimensional amplitude $a_0$ is half the wave
height (\cf{} \cref{sec:nondim}).
Note that the unforced KdV equation also has periodic solutions
known as cnoidal waves.
For a fixed height, these cnoidal waves have a smaller characteristic
wavelength $1/k_E$ than solitary waves and can be studied by choosing
larger $\mu_E > 6 \epsilon$ (\cf{} \cref{sec:comparison}).
However, wind-induced shape changes are more readily understood when
considering solitary waves owing to their reduced number of free
parameters (\ie{} $\mu_E$).
Furthermore, since solitary waves are well understood and highly
relevant to fluid dynamical
systems~\citep[\eg][]{hammack1974korteweg,miles1979korteweg,lin1998numerical,monaghan1999solitary},
we will restrict our analysis to solitary waves for brevity and clarity.
The wind-forcing term $P \partial_x^2 \eta_0$ in \cref{eq:kdv_burgers}
disrupts the solitary wave's balance of dispersion and nonlinearity,
inducing growth/decay and shape changes.
The KdV--Burgers equation has no known solitary-wave solutions, so we
will solve it numerically.

\subsection{\label{sec:numerics} Numerics and shape statistics}
\noindent
To solve \cref{eq:kdv_burgers} numerically, we will use the Dedalus
spectral solver~\citep{burns2020dedalus} which implements a generalized
tau method with a Chebyshev basis.
Since the onshore wind, $P>0$ case is ill posed, we require an implicit
solver, so time stepping is done with coupled four-stage, third-order
Diagonally Implicit Runge--Kutta and Explicit Runge--Kutta schemes.
The spatial domain has a length of $L=80$, and we require $\eta_0 = 0$
at $x' = -40$ and $\eta_0 = \partial_x \eta_0 = 0$ at $x' = 40$.
We employ $N_c = 1600$ Chebyshev coefficients and zero padding with a
scaling factor of $3/2$ to prevent aliasing of nonlinear terms.
This corresponds to $N_x = 2400$ spatial points with spacing $\Delta x =
\numrange{7.7e-5}{7.9e-2}$ for an average spacing of $\Delta x = 0.05$.
The simulation runs from $t_1 = 0$ to $t_1 = T = 10$, since the
multiple-scale expansion of \cref{sec:boussinesq} is only accurate for
times of $\order{1/\epsilon}$.
Adaptive time stepping is employed such that the
Courant--Friedrichs--Lewy number is $(\Delta t) \max(\eta_0)/(\Delta x)
= 1$.
For the unforced case, this corresponds to $\Delta t \approx
\num{7.86e-3}$, increases to \num{1.04e-2} for $P=-0.25$ and
decreases to \num{4.73e-4} for $P = 0.25$.
We found that linearly ramping up $P$ from $0$ at $t_1=0$ to its full
value at $t_1 = \epsilon$, or full, dimensional time $T_0 = 1/(\sqrt{gh}
k_E)$ (\ie{} the time required to cross the inverse, effective
wavenumber $1/k_E$, or `wave-crossing time') did not qualitatively
modify the results, so we do not utilize such a ramp-up here.
The spectral solver results in high numerical accuracy, with the
normalized root-mean-square difference between the unforced ($P=0$)
profile $\eta_0$ at $t'_1=10$ and the initial condition $\eta^{(0)}_0$
is \num{2e-13}, and the normalized wave height change is
$1-\bqty{\max(\eta_0) - \min(\eta_0)}/\allowbreak
\bqty{\max(\eta^{(0)}_0) - \min(\eta^{(0)}_0)} = \num{-1e-13}$.

We quantify the wave shape with the wave's energy $E$, skewness $\Sk$
and asymmetry $\As$,
\begin{subequations}
\begin{equation}
  E \coloneqq \langle \eta_0^2 \rangle , \quad
  \Sk \coloneqq \frac{\langle \eta_0^3 \rangle}{\langle \eta_0^2
  \rangle^{3/2}} \
  \qq{and}
  \As \coloneqq \frac{\langle \hilbert \Bqty{\eta_0^3} \rangle}{\langle
    \eta_0^2 \rangle^{3/2}}
  , \qq{with}
  \langle f \rangle \coloneqq \frac{1}{L} \int_{-L/2}^{L/2} f
  \dd{x} .
  \tag{\theequation\textit{a--c}}
  \label{eq:shape_stats_def}
\end{equation}
\end{subequations}
Here, $\hilbert(f)$ is the Hilbert transform of $f$, defined as the
imaginary part of $\fourier^{-1} \pqty{ \fourier \pqty{f} 2U}$ with $U$
the unit step function and $\fourier$ the Fourier transform.
Since these definitions depend on the domain size $L$, we normalize the
energy $E$ and skewness $\Sk$ by their initial values.

\section{\label{sec:results} Results}
\noindent
We study the pressure magnitude's effect on solitary-wave evolution and
shape by varying the KdV--Burgers equation's \cref{eq:kdv_burgers} one
free parameter, $P k_E/(\rho_w g \epsilon)$, with emphasis on the
contrast between onshore ($P > 0$) and offshore wind ($P < 0$).
We revert to denoting non-dimensional variables with primes and
dimensional ones without.

\begin{figure}
  \centering
  { 
    \phantomsubcaption{}\label{fig:snapshots_solitary:a}
    \phantomsubcaption{}\label{fig:snapshots_solitary:b}
  }
  \includegraphics{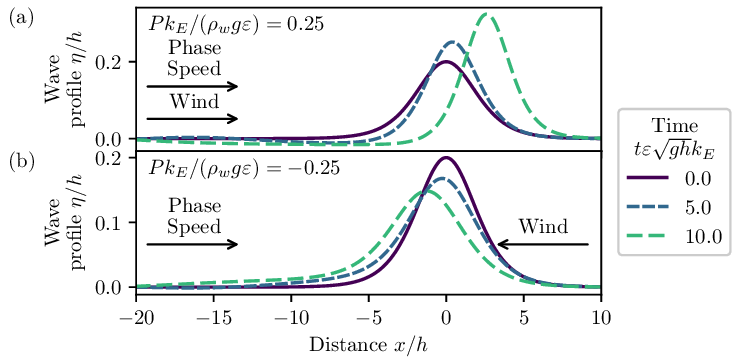}
  \caption{
    Solitary-wave evolution under
    \subref{fig:snapshots_solitary:a}
    onshore and
    \subref{fig:snapshots_solitary:b}
    offshore wind-induced surface pressure in the frame of the unforced
    solitary wave.
    Non-dimensional wave height $\eta/h$ versus
    non-dimensional distance $x/h$ for $\epsilon=0.1$,
    $\mu_E = 0.6$, $\abs{P k_E/(\rho_w g \epsilon)} = 0.25$ and
    non-dimensional slow times $t'_1 = t \epsilon \sqrt{gh} k_E = 0$,
    $5$ and $10$, as indicated in the legend.
    Only a subset of the full spatial domain is shown.
    The arrows denote the wave propagation (phase speed) and wind
    direction.
  }\label{fig:snapshots_solitary}
\end{figure}

The wave profile $\eta/h$ snapshots in \cref{fig:snapshots_solitary}
qualitatively show how the wave shape evolves over non-dimensional slow
time $t'_1 = t \epsilon \sqrt{g h} k_E$ in the unforced solitary wave's
frame.
The onshore wind generates wave growth, apparent at the wave
crest (\cref{fig:snapshots_solitary:a}), whereas the offshore wind causes
decay (\cref{fig:snapshots_solitary:b}).
The wind also changes the phase speed, with the wave's acceleration
(deceleration) under an onshore (offshore) wind visible by the advancing
(receding) of the crest.
This is expected due to the (unforced) solitary wave's
nonlinear phase speed \cref{eq:nonlin_phase_speed} dependence on the
wave height $H$.

In shallow water, wave growth/decay and phase speed changes are
well-known wind
effects~\citep[\eg][]{miles1957generation,cavaleri1981wind}, but
wind-induced wave shape changes~\citep{zdyrski2020wind} have not been
previously studied for shallow-water systems.
Such changes are visible in \cref{fig:snapshots_solitary} where, despite
the wave starting from a symmetric, solitary-wave initial condition, the
wind induces a horizontal asymmetry in the wave shape, particularly on
the rear face ($x<0$) of the wave.
The offshore wind (\cref{fig:snapshots_solitary:b}) raises the
rear base of the wave (near $x/h = -5$) relative to its initial profile
(purple line), but the onshore wind (\cref{fig:snapshots_solitary:a})
depresses the rear face and forms a small depression below the still
water level at $t\epsilon \sqrt{gh} k_E=5$ (blue line) which widens and
deepens at $t\epsilon \sqrt{gh} k_E=10$ (green line).
Finally, the onshore wind (\cref{fig:snapshots_solitary:a})
increases the maximum wave-slope magnitude with time while the offshore
wind (\cref{fig:snapshots_solitary:b}) decreases it, although the
windward side of the wave becomes steeper than the leeward side for both
winds (up to \SI{8}{\percent} steeper for the time period shown).
Although the equation is ill posed in the sense of Hadamard, the smooth
solutions show that our solution is acceptable up to the current time
and thus we are justified in neglecting a regularization scheme.

\begin{figure}
  \centering
  { 
    \phantomsubcaption{}\label{fig:snapshots_solitary_tail:a}
    \phantomsubcaption{}\label{fig:snapshots_solitary_tail:b}
  }
  \includegraphics{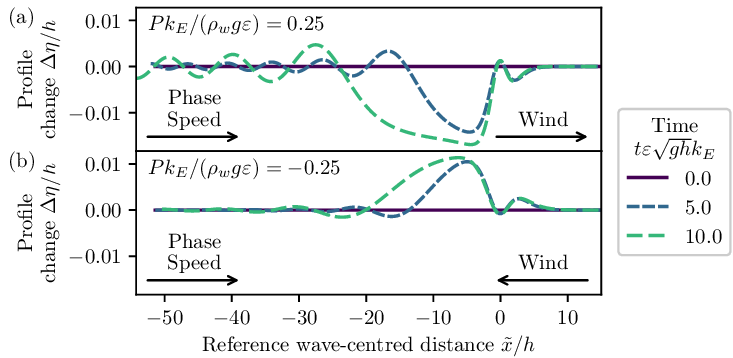}
  \caption{
    The non-dimensional profile change $\Delta \eta/h$ between the
    surface profile and reference solitary wave
    \cref{eq:initial_condition} under
    \subref{fig:snapshots_solitary_tail:a} onshore and
    \subref{fig:snapshots_solitary_tail:b} offshore Jeffreys forcing
    versus non-dimensional reference wave-centred distance
    $\tilde{x}/h$.
    Results are shown for $\epsilon=0.1$, $\mu_E = 0.6$, $\abs{P
    k_E/(\rho_w g \epsilon)} = 0.25$ and non-dimensional slow times
    $t'_1 = t \epsilon \sqrt{gh} k_E = 0$, $5$ and $10$, as indicated
    in the legend.
    Only a subset of the full spatial domain is shown.
    The arrows denote the direction of wave propagation (phase speed) or
    wind direction.
  }\label{fig:snapshots_solitary_tail}
\end{figure}

To further examine the wind-induced wave asymmetry, we fit $\eta$ to a
reference solitary-wave profile $\eta_{\text{ref}}$
\cref{eq:initial_condition} by minimizing the $L_1$ difference, yielding
the reference height $H_{\text{ref}}(t_1)$ and peak location
$x_{\text{ref}}(t_1)$.
The profile change is defined as $\Delta \eta(x) \coloneqq \eta -
\eta_{\text{ref}}$ and is shown as a function of the reference
wave-centred distance $\tilde{x} \coloneqq x - x_{\text{ref}}$ in
\cref{fig:snapshots_solitary_tail}.
Notice that the profile change begins near the front face of the wave
and has extrema for negative $\tilde{x}'$ but with opposite signs for
onshore and offshore winds.
Additionally, the magnitude of the extrema decay with distance in the
$-\tilde{x}$ direction.
Finally, note that the onshore (offshore) wind generates a small peak
(trough) at $\tilde{x} = 0$ and two small troughs (peaks) near
$\tilde{x}/h = \pm 3$, with the $\tilde{x}<0$ extrema larger than the
$\tilde{x}>0$ one.
This is analogous to a dispersive tail, well known in KdV-type
systems~\citep[\eg][]{hammack1974korteweg}, and its appearance here
helps explain the pressure-induced shape change (\cf{}
\cref{sec:physical_reason}).

\begin{figure}
  \centering
  { 
    \phantomsubcaption{}\label{fig:statistics_solitary:a}
    \phantomsubcaption{}\label{fig:statistics_solitary:b}
    \phantomsubcaption{}\label{fig:statistics_solitary:c}
  }
  \includegraphics{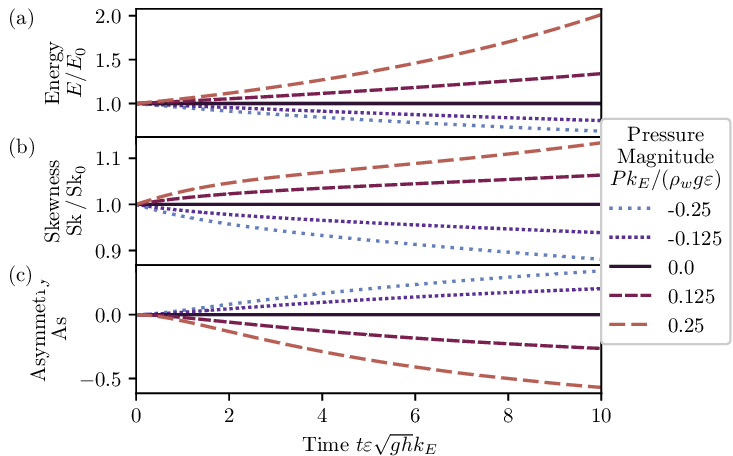}
  \caption{
    Solitary-wave shape statistics under onshore and offshore
    Jeffreys forcing versus non-dimensional slow time $t'_1 = t
    \epsilon \sqrt{gh} k_E = \numrange[range-phrase={\text{--}}]{0}{10}$.
    The
    \subref{fig:statistics_solitary:a}
    energy (normalized by the initial energy),
    \subref{fig:statistics_solitary:b}
    skewness (normalized by the initial skewness) and
    \subref{fig:statistics_solitary:c}
    asymmetry are defined in
    \cref{eq:shape_stats_def}.
    Results are shown for $\epsilon=0.1$, $\mu_E = 0.6$ and pressure
    magnitude $\abs{P k_E/(\rho_w g \epsilon)}$ up to $0.25$, as
    indicated in the legend.
    The solid black line is the unforced case, $P = 0$, and
    shows no growth or asymmetry and a constant skewness.
  }\label{fig:statistics_solitary}
\end{figure}

The effect of wind on wave shape is quantified by the time evolution of
wave shape statistics --- energy, skewness and asymmetry --- for onshore
and offshore wind~(\cref{fig:statistics_solitary}).
We plot all cases for initial steepness $\epsilon = 0.1$ up to slow time
$t \epsilon \sqrt{g h} k_E = 10$, corresponding to $10/\epsilon = 100$
wave-crossing times, $T_0 = 1/(\sqrt{gh} k_E)$.
The unforced case ($P=0$) displays constant shape statistics and zero
asymmetry, as expected.
The normalized energy $E/E_0$ shows different growth/decay rates:
the onshore wind ($P>0$) causes accelerating wave growth while the
offshore wind ($P<0$) causes slowing wave decay
(\cref{fig:statistics_solitary:a}).
The energy of the unforced wave is virtually unchanged, with a
normalized energy change of $1-E/E_0 = \num{-1e-13}$ at $t'_1 = 10$.
The onshore (offshore) wind causes the wave to become more (less) skewed
over time, with the normalized skewness nearly symmetric about unity
with respect to $\pm P$.
Finally, the onshore wind causes a backwards tilt and negative asymmetry
while the offshore wind increases the asymmetry and causes a forward
tilt, which was also seen in \cref{fig:snapshots_solitary}.
Notice that $\abs{\As}$ is larger for onshore winds than offshore winds.
Since the definitions of the skewness and asymmetry are insensitive to
waveform scaling $\eta \to \lambda \eta$, this effect is not simply
caused by the wave's growth/decay.
Instead, the onshore wind generates a larger dispersive tail
(\cref{fig:snapshots_solitary_tail}), which is the asymmetric wave
component.

\section{\label{sec:discussion} Discussion}

\subsection{\label{sec:press_mag} Wind speed estimation}
\noindent
We now relate the non-dimensional pressure magnitude $P k_E/(\rho_w g) =
\order{\epsilon}$ to the wind speed.
First, we need a relationship between the surface pressure and wave
energy $E$ \cref{eq:shape_stats_def}, which we can approximate using the
standard procedure \citep[\eg][]{mei2005nonlinear} of multiplying the
(non-dimensional, denoted by primes) KdV--Burgers equation
\cref{eq:kdv_burgers} by $\eta'_0$ and integrating from $x'=-\infty$ to
$\infty$ to obtain
\begin{equation}
  \pdv{t'_1} \int_{-\infty}^{\infty} {\eta'}_0^2 \dd{x'}
  = \int_{-\infty}^{\infty} P' \pqty{\pdv{\eta'_0}{x'}}^2
  \dd{x'} .
  \label{eq:energy_nondim}
\end{equation}
The left integral is the non-dimensional energy
\cref{eq:shape_stats_def}, so re-dimensionalizing and converting back to
the full time $t$ gives the energy growth rate $\gamma$,
\begin{equation}
  \frac{\gamma}{c_0 k_E} \coloneqq
  \frac{1}{c_0 k_E E} \pdv{E}{t}
  = \frac{P k_E}{\rho_w g} \frac{\langle (\partial_x \eta)^2 \rangle}
    {\langle (k_E \eta)^2 \rangle}
  = \frac{1}{5} \frac{P k_E}{\rho_w g}
  ,
  \label{eq:gamma_vs_P_solitary}
\end{equation}
with $\langle (\partial_x \eta)^2 \rangle / \langle (k_E \eta)^2 \rangle
= 1/5$ evaluated with the initial, solitary-wave profile
\cref{eq:initial_condition} and the linear, shallow-water phase speed
$c_0 = \sqrt{gh}$ coming from the re-dimensionalization of $t' = t c_0
k_E$ \cref{eq:nondim_expressions}.
Alternatively, a secondary multiple-scale approximation of the
forward KdV--Burgers equation has been used previously to derive
the energy growth rate for solitary waves as~\citep{zdyrski2019effects}
\begin{equation}
  E \propto \frac{1}{\pqty{1 - \gamma t}^2}
  \qq{with}
  \gamma \coloneqq b \bqty{\frac{P k_E}{\rho_w g}} c_0 k_E
  ,
  \label{eq:actual_energy}
\end{equation}
with analytically derived $b = 2/15$.
Numerically fitting \cref{eq:actual_energy} to our calculated energy
instead yields $b = \num{0.10081 \pm 0.00003}$, similar to the analytic
approximation.
Note that the exponential energy growth \cref{eq:gamma_vs_P_solitary}
correctly approximates \cref{eq:actual_energy} for small times $\gamma t
\ll 1$, and both expressions are consistent with the observed
accelerating (decelerating) energy change for $P>0$ ($P<0$) in
\cref{fig:statistics_solitary}.

Next, \citeauthor{jeffreys1925formation}'s
\citeyearpar{jeffreys1925formation} theory relates the growth rate of
periodic waves to the wind speed $U_{\lambda/2}$,
measured at a height equal to half the wavelength $z=\lambda/2$, as
\begin{equation}
  \frac{\gamma}{c k} = S_{\lambda/2} \frac{\rho_a}{\rho_w}
    \pqty{\frac{U_{\lambda/2}}{c}-1}
    \abs{\frac{U_{\lambda/2}}{c}-1} ,
  \label{eq:gamma_vs_u_jeffreys}
\end{equation}
with $S_{\lambda/2}$ a small, non-dimensional sheltering parameter
potentially dependent on $\epsilon$, $\mu_E$ and $U_{\lambda/2}/c$.
For simplicity, we approximate the nonlinear phase speed $c$ (given
non-dimensionally in \cref{eq:nonlin_phase_speed}) by its leading-order
term $c_0 = \sqrt{gh}$, yielding an error of only \SI{10}{\percent} in
the subsequent calculations.
Combining this approximation of \cref{eq:gamma_vs_u_jeffreys} with
\cref{eq:gamma_vs_P_solitary} gives
\begin{equation}
  U_{\lambda/2} = c_0 \pqty{1 \pm \sqrt{\frac{1}{5} \abs{\frac{P k_E}{\rho_w g}}
    \frac{\rho_w}{\rho_a} \frac{1}{S_{\lambda/2}}}} .
  \label{eq:U_vs_P}
\end{equation}
Here, the $\pm$ corresponds to onshore ($+$) or offshore ($-$) winds.
Note that changing the wind direction (\ie{} $\pm$ sign) while holding
the surface pressure magnitude $\abs{Pk_E/(\rho_w g)}$ constant means
onshore wind speeds $\abs{U_{\lambda/2}}$ will be larger than offshore
wind speeds.

We can evaluate \cref{eq:U_vs_P} for the parameters of
\cref{sec:results}: $\epsilon=0.1$, $\mu_E = 0.6$ and $Pk/(\rho_w g
\epsilon) = 0.25$.
\Citet{donelan2006wave} parameterized $S_{\lambda/2}$ for
periodic shallow-water waves with a dependence on airflow separation:
$S_{\lambda/2} = 4.91 \epsilon \sqrt{\mu}$ for our non-separated
flow (according to their criterion), with $\mu \coloneqq (kh)^2$.
Assuming this holds approximately for solitary waves, we choose $\lambda
= 2 \pi/k_E = \SI{20}{\meter}$ to calculate the wind speed at $z
= \lambda/2 = \SI{10}{\meter}$.
This choice corresponds to a depth of $h = \SI{2.5}{\meter}$ and initial
wave height $H_0 = \SI{0.5}{\meter}$ and yields a wind speed of $U_{10}
= \SI{22}{\meter\per\second}$, a physically realistic wind speed for
strongly forced shallow-water waves.
Weaker wind speeds will induce smaller surface pressures and thus take
longer to change the wave shape.

\subsection{\label{sec:physical_reason} Physical mechanism of asymmetry
generation}
\noindent
Our initial, symmetric solitary waves \cref{eq:initial_condition} are
permanent-form solutions of the unforced KdV equation.
More generally, any initial solitary wave which does not exactly solve
the KdV equation will evolve into a solitary wave and a trailing,
dispersive tail according to the inverse scattering
transform~\citep[\eg][]{mei2005nonlinear}.
In our system, the pressure continually perturbs the system away from
the unforced KdV soliton solution resulting in a trailing, bound,
dispersive tail (\cref{fig:snapshots_solitary_tail}), which is
responsible for the wave asymmetry.
To see this, consider an initial, symmetric profile $\eta$.
The pressure forcing term $P \partial_x^2 \eta$ preserves the initial
symmetry and induces a symmetric bound wave after a short time $\Delta
t'_1 \ll 1$.
This is apparent when considering the non-dimensional KdV--Burgers
equation \cref{eq:kdv_burgers} in the unforced solitary wave's frame
(\cref{fig:snapshots_solitary}) at the initial time,
\begin{align}
  &\eval{\pdv{\eta'_0}{t_1}}_{t'_1=0} = -P' \pdv[2]{x} \bqty{
  \sech^2\pqty{\frac{x'}{2}}}
  \\
  &\quad \implies \eta'_0(x', \Delta t'_1) =
  (2-P'\Delta t'_1) \sech^2\pqty{\frac{x'}{2}}
  +
  P' \Delta t'_1 \frac{3}{2}\sech^4\pqty{\frac{x'}{2}}
  .
  \label{eq:pressure_change}
\end{align}
The $P'\Delta t'_1$ terms generate a small bound wave with a peak
(trough) at $x'=0$ and troughs (peaks) symmetrically in front and behind
the wave peak for onshore (offshore) wind.
As time increases, the continual pressure forcing causes the bound wave
to grow and lengthen behind the wave, as is apparent in
\cref{fig:snapshots_solitary_tail} (\eg $\tilde{x}/h =
\numrange{-20}{3}$ for $P'=0.25$ and $t'_1=10$).

The small numerical value $\abs{P'} = 0.25 \ll 1$ used in
\cref{sec:results} allows us to consider the wave's evolution as two
steps with time scale separation.
First, the pressure generates a bound wave \cref{eq:pressure_change} on
the slow time scale, and then the wave evolves a dispersive tail on the
fast time scale according to the inverse scattering transform of the
unforced KdV equation.
The dispersive tail in \cref{fig:snapshots_solitary_tail} (\eg located
left of $\tilde{x}/h=-20$ for $P'=0.25$ and $t'_1=10$) is analogous to the
ubiquitous dispersive tails in prior studies on shallow-water solitary
waves, such as \figsname{} 8(b) and 8(c) of \citet{hammack1974korteweg}.
However, unlike dispersive tails generated from initial conditions which
fail to satisfy the KdV equation, our tail is continually forced and
lengthened by the wind forcing.
Finally, interactions with the trailing, dispersive tail are responsible
for lengthening the bound wave \cref{eq:pressure_change} behind, rather
than ahead, of the solitary wave.
Hence, the disturbance induced by the pressure forcing
\cref{eq:pressure_change} has two effects on the wave.
First, the wind slowly generates a bound wave which changes the height
and width of the initial solitary wave, which is reflected in the growth
(decay) and narrowing (widening) under onshore (offshore) winds in
\cref{fig:snapshots_solitary}.
Second, it quickly generates an asymmetric, dispersive tail behind the
wave (\cref{fig:snapshots_solitary_tail}), producing a greater shape
change on the wave's rear face (\cref{fig:snapshots_solitary}).
Finally, the different wind directions (\ie{} pressure forcing signs)
change the sign of the bound wave and dispersive tail and, hence, the
sign of the asymmetry in \cref{fig:statistics_solitary}.

\subsection{\label{sec:comparison} Comparison to intermediate and deep water}
\noindent
\Citet{zdyrski2020wind} investigated the effect of
wind on Stokes-like waves in intermediate to deep water.
This study, with wind coupled to waves in shallow water, finds
qualitative agreement with those intermediate- and deep-water results.
The shallow-water asymmetry magnitude increases as the pressure
magnitude $P$ increases~(\cref{fig:statistics_solitary}), and \figname{}
4(a) of \citet{zdyrski2020wind} displayed a similar trend for the
corresponding Jeffreys pressure profile, with positive (negative)
pressure increasing (decreasing) the asymmetry.
Although \citet{zdyrski2020wind} compared their theoretical
predictions to limited experimental results with $kh > 1$, there are no
appropriate experiments on wind-induced changes to wave shape in shallow
water for comparison with our results.
In addition to the Jeffreys pressure profile employed here,
\citet{zdyrski2020wind} also utilized a generalized Miles profile,
only applicable to periodic waves, wherein the pressure was proportional
to $\eta$ shifted by a distance parameter $\psi_P/k$.
Future investigations could couple a higher-order Zakharov
equation~\citep[\eg][]{dommermuth1987high} to a Jeffreys-type pressure
forcing or to an atmospheric large eddy simulation, as was done for deep
water by \citet{hao2019wind}.
Although this analysis focuses on solitary waves, we also investigated
the effect of wind on periodic waves using the cnoidal-wave KdV
solutions as initial conditions.
Wind-forced cnoidal waves displayed qualitatively similar shape changes
with stronger onshore (offshore) wind causing the energy and skewness to
increase (decrease) while the asymmetry decreased (increased) with time.
Furthermore, results were qualitatively similar across multiple classes
of cnoidal waves with different values of $\mu_E$, implying that these
results apply rather generally.

\section{\label{sec:conclusion} Conclusion}
\noindent
Prior results~\citep{zdyrski2020wind} in intermediate and deep water
demonstrated that wind,
acting through a wave-dependent surface pressure,
can generate shape changes that become more pronounced in
shallower water.
Here, we produced a novel analysis of wind-induced wave shape changes in
shallow water using a multiple-scale analysis to couple weak wind with
small, shallow-water waves, \ie{} $a_0/h \sim (k_E h)^2 \sim P k/(\rho_w
g) \ll 1$.
This analysis produced a KdV--Burgers equation governing the wave
profile $\eta$, which we then solved numerically with a symmetric,
solitary-wave initial condition.
The deviations between the numerical results and a reference solitary
wave had the form of a bound, dispersive tail, with differing signs for
onshore and offshore wind.
The tail's presence and shape are the result of a symmetric,
pressure-induced shape change evolving under the inverse scattering
transform.
We also estimated the energy, skewness and
asymmetry as functions of time and pressure magnitude.
For onshore wind (positive $P$), the wave's energy and skewness
increased with time while asymmetry decreased, while offshore wind
produced the opposite effects.
Furthermore, these effects were enhanced for strong pressures, and they
reduced to the unforced case for $P=0$.
The shape statistics found here show qualitative agreement with the
results in intermediate and deep water.
Finally, the wind speeds corresponding to these pressure differences
were calculated and found to be physically realistic.

\begin{acknowledgements}
\noindent
\textbf{Acknowledgements.}
We are grateful to D.G.~Grimes and M.S.~Spydell for discussions on
this work.
Additionally, we thank the anonymous reviewers for their suggestions and
comments.

\vspace{\baselineskip}
\noindent
\textbf{Funding.}
We thank the National Science Foundation (OCE-1558695) and the Mark Walk
Wolfinger Surfzone Processes Research Fund for their support of this
work.

\vspace{\baselineskip}
\noindent
\textbf{Declaration of interests.}
The authors report no conflict of interest.
\end{acknowledgements}

\bibliographystyle{jfm-abbrv-3}
\bibliography{references}

\end{document}